\documentclass[pdflatex,sn-mathphys-ay,oneside,iicol]{sn-jnl-qmi}

\usepackage{graphicx}%
\usepackage{multirow}%
\usepackage{amsmath,amssymb,amsfonts}%
\usepackage{amsthm}%
\usepackage{mathrsfs}%
\usepackage[title]{appendix}%
\usepackage{xcolor}%
\usepackage{textcomp}%
\usepackage{manyfoot}%
\usepackage{booktabs}%
\usepackage{algorithm}%
\usepackage{algorithmicx}%
\usepackage{algpseudocode}%
\usepackage{listings}%
\usepackage{comment}

\usepackage{lmodern}
\usepackage{anyfontsize}
\usepackage{braket}
\usepackage{bbold}
\usepackage[nolist,nohyperlinks]{acronym}
\usepackage[capitalize,noabbrev]{cleveref}

\begin{document}

\acrodef{pqc}[PQC]{parametrized quantum circuit}
\acrodef{rl}[RL]{reinforcement learning}
\acrodef{qrl}[QRL]{quantum reinforcement learning}
\acrodef{qml}[QML]{quantum machine learning}
\acrodef{qpg}[QPG]{quantum policy gradients}
\acrodef{regqpg}[RegQPG]{regularized quantum policy gradients}
\acrodef{cregqpg}[CurrRegQPG]{curriculum regularized quantum policy gradients}

\title{Robustness and Generalization in Quantum Reinforcement Learning via Lipschitz Regularization}


\author*[1,2]{\fnm{Nico} \sur{Meyer}}\email{nico.meyer@iis.fraunhofer.de}
\author[3]{\fnm{Julian} \sur{Berberich}}
\author[1]{\fnm{Christopher} \sur{Mutschler}}
\author[1]{\fnm{Daniel D.} \sur{Scherer}}

\affil[1]{\orgdiv{Fraunhofer IIS}, \orgname{Fraunhofer Institute for Integrated Circuit IIS}, \orgaddress{\city{N\"urnberg}, \country{Germany}}}
\affil[2]{\orgdiv{Pattern Recognition Lab}, \orgname{Friedrich-Alexander-Universit\"at Erlangen-N\"urnberg}, \orgaddress{\city{Erlangen}, \country{Germany}}}
\affil[3]{\orgname{University of Stuttgart}, \orgdiv{Institute for Systems Theory and Automatic Control}, \orgaddress{\city{Stuttgart}, \country{Germany}}}

\abstract{
Quantum machine learning leverages quantum computing to enhance accuracy and reduce model complexity compared to classical approaches, promising significant advancements in various fields. Within this domain, quantum reinforcement learning has garnered attention, often realized using variational quantum circuits to approximate the policy function. This paper addresses the robustness and generalization of quantum reinforcement learning by combining principles from quantum computing and control theory. Leveraging recent results on robust quantum machine learning, we utilize Lipschitz bounds to propose a regularized version of a quantum policy gradient approach, named the RegQPG algorithm. We show that training with RegQPG improves the robustness and generalization of the resulting policies. Furthermore, we introduce an algorithmic variant that incorporates curriculum learning, which minimizes failures during training. Our findings are validated through numerical experiments, demonstrating the practical benefits of our approach.
}


\maketitle

\section{\label{sec:introduction}Introduction}

\Ac{qml} has emerged as a promising field at the intersection of quantum computing and of machine learning~\citep{biamonte2017quantum}. It is conjectured to be a candidate for using noisy intermediate-scale quantum (NISQ) devices~\citep{preskill2018quantum} to realize quantum utility. Under specific conditions, it is known, that \ac{qml} can unveil patterns intractable for classical methods. This enables exponential enhancements in model accuracy and convergence speed~\citep{liu2021rigorous} for specific problem instances. With these advantages, QML is poised to revolutionize various domains, including optimization, cryptography, and drug discovery~\citep{zaman2023survey}.

Within the domain of \ac{qml}, \ac{qrl} has gained significant attention over recent years. This subfield lies at the intersection of quantum computing, see~\cite{nielsen2011quantum} for an introduction, and classical \ac{rl}, as outlined by~\cite{sutton2018reinforcement}. A comprehensive overview of \ac{qrl} can be found in~\cite{meyer2022survey}, highlighting some key challenges and approaches in the field. Often the central idea is to use a \ac{pqc}~\citep{cerezo2021variational} as a function approximator. Initially, there have been proposals to parameterize the Q-function~\citep{chen2020variational,skolik2022quantum}. Later papers extended the concept to direct parameterization of the policy~\citep{jerbi2021parametrized,meyer2023quantum}, followed up by more sophisticated algorithmic variants~\citep{wu2020quantum,yun2022quantum}. More recent work discusses offline \ac{qrl}~\citep{periyasamy2023bcqq}, and the use of variational LSE solvers~\citep{bravo2023variational,meyer2024comprehensive} for quantum policy iteration~\citep{meyer2024warm}.

When applying \ac{rl} algorithms to complex and safety-critical applications ensuring reliable behavior becomes paramount. 
A key challenge is to cope with practical limitations such as noisy measurements or limited data samples, complicating the generalization from training data to unseen scenarios.
In classical \ac{rl}, generalization has received significant attention~\citep{packer2019assessing}.
Common approaches employ, e.g., robust policies which are insensitive to environment changes~\citep{tamar2015optimizing} or adaptive policies which react to environment changes in a suitable way~\citep{yu2017preparing}.
In particular, the generalization capability of a policy can be quantified in terms of its Lipschitz continuity properties~\citep{wang2019generalization}.

Our paper contributes to the field of \ac{qrl} as follows: We modify the training procedure of \ac{qpg} with additional Lipschitz-based regularization, dubbed \ac{regqpg}. The underlying theory is derived in \cref{sec:training}, and the experimental setup used in this paper is described in \cref{sec:setup}. We formulate and quantify the robustness of the respective trained \ac{qrl} policies in \cref{sec:robustness}. Our findings indicate, that \ac{regqpg} enhances the robustness over a wide range of state perturbation strengths. Furthermore, we extend our analysis to consider generalization capabilities in \cref{sec:generalization}. The results in \cref{subsec:generalization_regqpg} show a clearly enhanced generalization performance over different ranges of initial conditions when using Lipschitz regularization. Moreover, the extension with concepts from curriculum learning in \cref{subsec:generalization_cregqpg} -- dubbed \ac{cregqpg} -- further improves these generalization capabilities, while also minimizing the number of failures during training. Finally, \cref{sec:discussion} discusses the significance of our findings for the \ac{qrl} community, proposes future extensions, and concludes the paper.


\section{Regularized QRL Training}\label{sec:training}

\begin{figure}[tbp]
    \centering
    \includegraphics[width=1.0\linewidth]{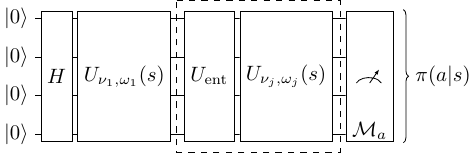}
    \caption{The Quantum model used as the policy approximator in the proposed \ac{regqpg} algorithm. The parametrized unitaries are composed of single-qubit rotations with trainable variational weights $\nu_j$ and encoding parameters $\omega_j$. This is followed by a static entanglement unitary and repeated several times. In the end, a tensored Pauli-Z observable is measured and plugged into \cref{eq:policy_postprocess} to approximate the policy value.}\label{fig:pqc}
\end{figure}

We consider a common \ac{rl} setup, where the action $a\in\mathcal{A}$ is determined via a parameterized policy 
\begin{align}
    a \sim \pi_{\Theta}(a | s),
\end{align}
with $\sum_a \pi_{\Theta}(a | s) = 1$.
Here, $s\in\mathcal{S}\subseteq\mathbb{R}^{n}$ denotes the environment state, $a \in \mathcal{A}$ is an action from a discrete set, and $\Theta$ summarizes the trainable parameters. We furthermore introduce the notion
\begin{align}\label{eq:pqc_vector}
    \pi_{\Theta}(\cdot | s) = \left[ \pi_{\Theta}(a=0|s), \pi_{\Theta}(a=1|s), ... \right],
\end{align}
mapping the environment state to the space of probability distributions over the actions.
Our goal is to iteratively adapt the parameters $\Theta$ to maximize the cumulative and discounted return
\begin{align}\label{eq:cost}
    J(\Theta)=\sum_{t'=t}^{T-1}\gamma^{t'-t} r_{t'}
\end{align}
for some horizon $T$ and discount factor $0 \leq \gamma \leq 1$. Note that the reward $r_t$ depends on the evolution of the state $s_t$ under the action sampled from $\pi_{\Theta}(s_t)$ as well as the action $a_t$ itself, which is why $J(\Theta)$ is a function of the parameters $\Theta$. Further details on the concept of \ac{rl} can be found in~\cite{sutton2018reinforcement}.

In this paper, we use a \ac{pqc} to approximate the policy $\pi_{\Theta}$. More specifically, $\pi_{\Theta}$ is represented by a quantum model with variational parameters, in particular including a trainable encoding~\citep{berberich2023training}. This can be expressed as
\begin{align}\label{eq:policy}
    \pi_{\Theta}(a|s)=\braket{0|U_{\Theta}(s)^\dagger\mathcal{P}_a U_{\Theta}(s)|0},
\end{align}
with some projector $\mathcal{P}_a$ and parametrized quantum circuit $U_{\Theta}(s)$. For the policy to form a probability distribution over the actions, we furthermore require that $\sum_a \mathcal{P}_a = \mathbb{1}$ and $\mathcal{P}_a \mathcal{P}_{a'} = \delta_{a,a'} \mathcal{P}_a$.
The latter is defined as
\begin{align}\label{eq:ansatz_1}
    U_{\Theta}(s)=U_{N,\Theta_N}(s)\cdots U_{1,\Theta_1}(s)
\end{align}
for layers $j=1,\dots,N$ and unitary operators 
\begin{align}\label{eq:ansatz_2}
    U_{j,\Theta_j}(s)=e^{-i(\nu_j + \omega_j^\top s)H_j},
\end{align}

More precisely, we train a policy that not only maximizes the return $J(\Theta)$ but also admits a small Lipschitz bound.
Mathematically, a Lipschitz bound of a function $f$ is any number $L>0$ such that 
\begin{align}\label{eq:lipschitz_bound_f_definition}
    \lVert f(x)-f(x')\rVert\leq L\lVert x-x'\rVert
\end{align}
holds for any $(x,x')$.
For each action $a$, a Lipschitz bound on the \ac{pqc}~\eqref{eq:policy} can be derived as
\begin{align}\label{eq:lipschitz_bound_a_k}
    L_{\Theta,a}=2\lVert
    \mathcal{P}_{a}\rVert\sum_{j=1}^N\lVert \omega_j\rVert\lVert H_j\rVert,
\end{align}
for details see~\cite{berberich2023training}.
This implies the following Lipschitz bound for the vector of probability distributions~\eqref{eq:pqc_vector}
\begin{align}\label{eq:lipschitz_bound}    L_{\Theta}=\sum_a L_{\Theta,a}=\sum_a 2\lVert
    \mathcal{P}_{a}\rVert\sum_{j=1}^N\lVert \omega_j\rVert\lVert H_j\rVert.
\end{align}
In the context of \ac{rl}, a small Lipschitz bound implies robustness of the policy against input perturbations, which may, e.g., arise due to noisy measurements of the state $s_t$ (see~\cref{sec:robustness});
further, a small Lipschitz bound enhances the generalization of the policy over varying initial conditions, both during and after training (see~\cref{sec:generalization}). Motivated by these facts, in this paper, we solve a multiobjective optimization problem:
maximize $J(\Theta)$ while keeping $L_{\Theta}$ small.
Note that the Lipschitz bound depends directly on the norm of the trainable parameters $\omega_j$.
Thus, we define a regularized reward function which encodes these two objectives
\begin{align}
    \label{eq:reg_loss}
    J_{\mathrm{reg}(\lambda)}(\Theta)=J(\Theta)-\lambda\sum_{j=1}^N\lVert \omega_j\rVert^2\lVert H_j\rVert^2,
\end{align}
where $\lambda\geq0$ is a hyperparameter which can be tuned by the user, e.g., via cross-validation.
Larger values of $\lambda$ encourage a more robust and stabilizing policy at the price of a possibly reduced reward $J(\Theta)$.
We train the policy via gradient ascent 
\begin{align}
    \label{eq:reg_update}
    \Theta\leftarrow\Theta+\alpha\nabla_{\Theta}J_{\mathrm{reg}(\lambda)}(\Theta),
\end{align}
resulting in a so-called \emph{policy gradient} approach~\citep{sutton1999policy}. The gradient of the regularized objective is obtained by
\begin{align} \label{eq:reg_grad}
    \nabla_{\Theta}J_{\mathrm{reg}(\lambda)}(\Theta)=\nabla_{\Theta}J(\Theta)-\nabla_{\Theta}\left(\lambda\sum_{j=1}^N\lVert \omega_j\rVert^2\lVert H_j\rVert^2\right),
\end{align}
where $\nabla_{\Theta}J(\Theta)$ can be estimated with the \emph{policy gradient theorem}~\citep{sutton1999policy}. This concept has already been generalized to \emph{quantum policy gradients} in previous work~\citep{jerbi2021parametrized,meyer2023quantum}. 
Note that the regularization causes an additional linear term in the update of the trainable encoding parameters $\omega_j$, i.e., 
\begin{align}
    w_j\leftarrow w_j+\alpha\nabla_{\omega_j}J(\Theta)-2\alpha\lambda\lVert H_j\rVert^2 \omega_j.
\end{align}
Since the Lipschitz bound~\eqref{eq:lipschitz_bound} is independent of the parameters $\nu_j$, the update of the \emph{standard} variational parameters $\nu_j$'s is not influenced by the regularization. Due to the main components of the proposed routine, we furthermore refer to it as the \acf{regqpg} algorithm. 
In the remainder of the paper, we study the impact of the above regularization on robustness (\cref{sec:robustness}) and generalization (\cref{sec:generalization}) of the trained policy (\cref{subsec:generalization_regqpg}). Furthermore, we develop a variant based on curriculum learning to further improve generalization and minimize the number of failures during training in (\cref{subsec:generalization_cregqpg}), dubbed the \ac{cregqpg} algorithm.


\section{Experimental Setup}\label{sec:setup}

\begin{figure}[tbp]
    \centering
    \includegraphics{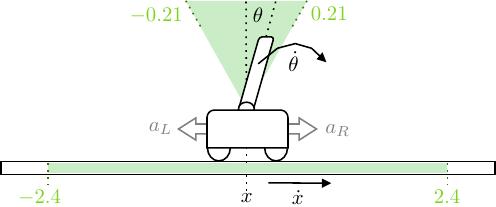}
    \caption{\label{fig:cartpole}Sketch of \texttt{CartPole} environment, see also~\cref{tab:cartpole} for details on observations.}
\end{figure}

\begin{table}[tbp]
    \centering
    \caption{Observations from the \texttt{CartPole} environment, refer also to \cref{fig:cartpole} for sketch. We report the admissible \emph{range} of observations, the corresponding \emph{initial} conditions, the factor we use to \emph{norm}alize to the approximate range $[-1, 1]$ for encoding, and the \emph{unit} in which the quantities are measured.}\label{tab:cartpole}%
    \begin{tabular*}{\linewidth}{@{}c|cccc@{}}
        \toprule
        & \multicolumn{2}{c}{Cart} & \multicolumn{2}{c}{Pole} \\
        \cmidrule{2-5}
        \cmidrule{4-5}
        & Position $x$ & Velocity $\dot{x}$ & Angle $\theta$ & Velocity $\dot{\theta}~$ \\
        \midrule
        Range & $~[-2.4, 2.4]~$ & $[-\infty, \infty]$ & $[-0.21, 0.21]$ & $[-\infty, \infty]$ \\
        Initial & \multicolumn{4}{c}{all uniform at random in range $[-0.05, 0.05]$\footnotemark[1]} \\
        Norm & $2.4$ & $2.5$ & $0.21$ & $2.5$ \\
        Unit & m & m/s & rad & rad/s \\
        \botrule
    \end{tabular*}
    \footnotetext[1]{Standard setting of initial conditions for \texttt{CartPole} environment, refers to unnormalized observations.}
\end{table}

In the following, we will describe the detailed algorithmic specifications and hyperparameter setting used for training the \ac{qrl} agents, and for all experiments in~\cref{sec:robustness,sec:generalization} building upon this. All implementations were realized using the \texttt{qiskit-torch-module}~\citep{meyer2024qiskit}, an extension of the Qiskit library~\citep{qiskit2024} for the fast training of quantum neural networks. A repository supporting full reproducibility of the results can be accessed as noted in the code availability statement.

\begin{figure}[tbp]
    \centering 
    \includegraphics{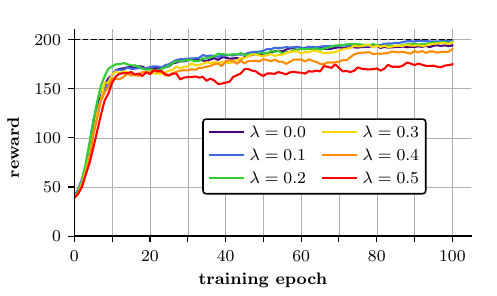}
\caption{\label{fig:train}Training using the \ac{regqpg} algorithm on the \texttt{CartPole} environment. One epoch performs updates on trajectories from $10$ environment instances. The strength of Lipschitz regularization is indicated by $\lambda$. Training performance is averaged over $100$ random seeds.}
\end{figure}

The used \ac{pqc} is a modified version of an ansatz frequently employed in \ac{qpg} algorithms~\citep{jerbi2021parametrized,meyer2023quantum,meyer2023natural} and is sketched in~\cref{fig:pqc}. It includes variational parameters $\nu$ and trainable encoding parameters $\omega$, satisfying~\cref{eq:ansatz_1,eq:ansatz_2}.
After an initially created equal superposition, parameterized unitaries $U_{\nu_j,\omega_j}(s)$ and all-to-all Pauli-Z entanglement unitaries $U_{\mathrm{ent}}$ are applied in alternation:
\begin{align}
    U_{\nu,\omega}(s) = U_{\nu_L,\omega_L}(s) \cdots U_{\mathrm{ent}} U_{\nu_1,\omega_1}(s) H^{\otimes n} \ket{\mathbf{0}}
\end{align}
Hereby, $L$ denotes the number of layers -- we use $L=3$ for all experiments. The individual parameterized unitaries are composed as
\begin{align} \label{eq:circuit}
    U_{\nu_j,\omega_j}(s)=U_{\nu_j}U_{\omega_j}(s),\>~j= 1,2,\cdots,L,
\end{align}
where we implement the state-independent and state-dependent parts individually using single-qubit rotations, i.e.,
\begin{align} \label{eq:circuit_1}
    U_{\nu_j}&=\bigotimes\nolimits_{i} R_y(\nu_{j,i,1}) R_z(\nu_{j,i,0}) \\ \label{eq:circuit_2}
    U_{\omega_j}(s)&=\bigotimes\nolimits_{i} R_z(\omega_{j,i,1} \cdot s_i) R_z(\omega_{j,i,0} \cdot s_i).
\end{align}
This formulation is an instance of the more general version in \cref{eq:ansatz_1,eq:ansatz_2}. This could be seen by replacing the $\omega_{j,i,k}$, with one-hot encoded vectors, i.e., $[0, \cdots, w_{j,i,k}, \cdots, 0]^t s = w_{j, i, k} \cdot s_i$, with $k=0, 1$. In the context of (quantum) policy gradients, we employ a stochastic policy \ $\pi(a|s) \mapsto [0,1]$, with $\sum_a \pi(a|s)=1$. For two actions $a=\left\{ 0, 1\right\}$, we can use the explicit definition
\begin{align}
    \label{eq:policy_postprocess}
    \pi(a|s) = \frac{(-1)^a \cdot \braket{U_{\nu,\omega}(s)^\dagger \mid Z^{\otimes n} \mid U_{\nu,\omega}(s)}+1}{2},
\end{align}
which by Born's rule is guaranteed to be a valid probability density function. There also exist extensions to larger action spaces~\citep{meyer2023quantum}.

For experimentally validating the claims in this paper, we use the \texttt{CartPole} environment~\citep{brockman2016openai}, sketched in~\cref{fig:cartpole}. The task is to
balance a pole mounted on a car by steering it left or right while avoiding violating spatial or angular conditions. Details on admissible values for the entries of the four-dimensional state are summarized in~\cref{tab:cartpole}. For encoding into the \ac{pqc}, the entries are normalized to the range $\left[ -1, 1 \right]$, where the normalization factors for velocity $\dot{x}$ and angular velocity $\dot{\theta}$ hold for typically encountered values. A reward of $r=1$ is given for each successful timestep, with a horizon of $T=200$. During training, the rewards are discounted with a factor $\gamma=0.99$ following~\cref{eq:cost}. As the state is four-dimensional, an instance with $n=4$ qubits of the ansatz defined in~\cref{eq:circuit,eq:circuit_1,eq:circuit_2} is used.

We trained with the described setup for $100$ epochs, each performing updates on a batch of $10$ full trajectories, using a learning rate of $\alpha=0.05$. The results for different regularization rates $\lambda$ -- following the procedure described in~\cref{sec:training} -- are shown in~\cref{fig:train}. All curves are averaged over $100$ random seeds. The training converges to a near-optimal policy in the absence of regularization (i.e., $\lambda=0.0$), and for regularization rates $\lambda=\{0.1;0.2;0.3$\}, with no significant differences in convergence behavior. For a higher regularization rate of $\lambda=0.4$, and especially $\lambda=0.5$, the procedure does not converge to the optimal reward of $200$, since the regularization outweighs the reward, compare~\cref{sec:training}.
In \cref{sec:robustness} and \cref{sec:generalization}, we demonstrate that the proposed regularization significantly improves the robustness and generalization of the learned policy.


\section{Robustness of QRL Policies}\label{sec:robustness}

\begin{figure}[tbp]
    \centering
    \includegraphics{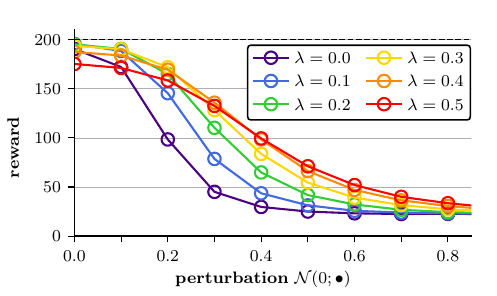}
    \caption{\label{fig:robustness}Lipschitz regularization enhances robustness of policies trained with the \ac{regqpg} algorithm. The models from \cref{fig:train} are tested on modified \texttt{CartPole} environments with observation perturbation, i.e.\ additive zero-mean Gaussian noise with increasing variance. Testing performance is averaged over 100 random seeds for each model, resulting in 10000 samples contributing to each data point.}
\end{figure}

By definition, a Lipschitz bound of the policy $\pi_{\Theta}$ quantifies robustness against perturbed state measurements. To be precise, suppose that instead of the true state $s$ only a perturbed measurement $\tilde{s}=s+\varepsilon$ is available, e.g., due to measurement noise or an adversarial attack~\citep{goodfellow2014explaining,szegedy2014intriguing}. Then, due to~\eqref{eq:lipschitz_bound_f_definition}, the change of the policy $\pi_{\Theta}$ is bounded as 
\begin{align} \label{eq:robustness}
    \lVert\pi_{\Theta}(\cdot|\tilde{s})-\pi_{\Theta}(\cdot|s)\rVert 
    \leq L_{\Theta}\lVert\varepsilon\rVert.
\end{align}
Thus, a smaller Lipschitz bound $L_{\Theta}$ implies an improved worst-case robustness against (additive) state perturbations.

In \cref{fig:robustness}, we demonstrate that the regularized training strategy proposed in \cref{sec:training} does indeed significantly improve the robustness of the trained policy. To this end, we assumed that the observations are perturbed with zero-mean Gaussian noise of increasing variance. The noise value is sampled from the respective distribution and applied to each element of the four-dimensional state individually. This happens after observation normalization as noted in~\cref{tab:cartpole}, i.e., the offset is applied to values in the approximate range $\left[ -1, 1 \right]$. It is important to note that the actual state of the environment is not affected by that perturbation.

We test all $100$ trained policies for each regularization rate from~\cref{sec:training} on $100$ instances of the described \texttt{Perturbated CartPole} environment, for perturbation strengths ranging from $\mathcal{N}(0;0.0)$ (i.e.\ no perturbation) to $\mathcal{N}(0;0.8)$. As expected, for each trained policy, the obtained reward decreases with increasing perturbation. In particular, the performance drops to a value of about $25$. This is the performance of a basically random agent on the \texttt{CartPole} environment, which can be explained by the observations being so noisy, that no informed decisions can be made. Further, while small regularization parameters $\lambda=\{0.0;0.1;0.2;0.3\}$ correspond to the same optimal performance obtained during training in the absence of noise, significant differences can be observed for increasing noise levels. In particular, \cref{fig:robustness} shows a direct correlation between higher regularization parameters $\lambda$ (i.e., smaller Lipschitz bounds of the trained policies) and higher reward for non-trivial noise levels. On the other hand, too high regularization, e.g., $\lambda=~0.5$, leads to a substantially lower reward for small noise levels since the regularized reward function~\eqref{eq:reg_loss} is unnecessarily biased towards overly robust solutions.

Notably, a sweet spot can be observed for regularization with $\lambda=0.3$, for which a substantially higher reward can be obtained than in the case without regularization (i.e.\ $\lambda=0.0$) across all considered noise levels. Overall, we can conclude, that Lipschitz regularization, as implemented in \ac{regqpg}, can significantly improve the robustness of quantum policies.


\section{Generalization of QRL Policies}\label{sec:generalization}

\begin{figure*}[tb]
    \centering
    \includegraphics{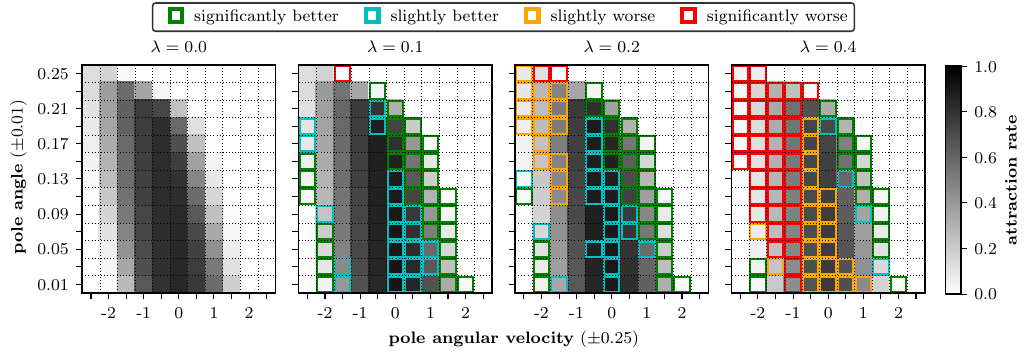}
    \caption{\label{fig:generalization_regqpg}Lipschitz regularization enhances generalization of policies trained with the \ac{regqpg} algorithm. During training, all features -- cart position, cart velocity, pole angle, pole angular velocity -- are by default initialized randomly at uniform in $\left[ -0.05, 0.05 \right]$. We test on a wider range of pole angles and angular velocities, and report the rate of attraction for these configurations, i.e.\ the fraction of successful test runs -- see also \cref{eq:rate_of_attraction}. The $100$ models for each regularization rate are evaluated for $100$ runs each, i.e.\ $10000$ samples contribute to each data point. Regions where the policies obtained via regularized training deviate from the non-regularized baseline and their corresponding full (half) confidence intervals -- defined by the variances -- do not overlap are highlighted as significantly (slightly) better or worse, respectively.}
\end{figure*}

In this section, we study the impact of Lipschitz regularization on the generalization of QRL policies. In particular, we show that Lipschitz regularization improves the ability of QRL policies to reliably control the underlying system even for initial states not included in the training data.
Intuitively, a small Lipschitz bound ensures that the policy $\pi_{\Theta}$ does not change too rapidly for varying states and, therefore, effects of overfitting are reduced.
While this connection is well-known in classical supervised machine learning~\citep{xu2012robustness} and has recently been explored in supervised QML~\citep{berberich2023training}, our results in this section exploit it for the first time in QRL.
In particular, in Section~\ref{subsec:generalization_regqpg}, we show that Lipschitz regularization indeed increases the region in which failure-free operation can be ensured after training.
Next, in Section~\ref{subsec:generalization_cregqpg}, we incorporate curriculum learning ideas for exploring the state space, which further improves generalization and minimizes the number of failures during training.


\subsection{Generalization of Trained Policies}\label{subsec:generalization_regqpg}

We experimentally support the claim that Lipschitz regularization enhances the generalization capabilities of trained policies in \cref{fig:generalization_regqpg}. For this, we evaluated the \emph{attraction rate} of the policies trained in \cref{sec:training} with regularization rates $\lambda=\{0.0;0.1;0.2;0.4\}$ -- for a definition see below. We found the pole angle to be the factor that much more frequently causes failed episodes, as opposed to the cart position. Therefore, we focus our analysis on the initial conditions for the pole, rather than the spatial factors. In particular, we evaluated initial conditions for the pole angular velocity sampled uniformly at random in $\left[ 0.00, 0.02 \right],~\left[ 0.02, 0.04 \right],~\dots,~\left[ 0.24, 0.26 \right]$, and pole angles sampled uniformly at random in $\left[ -2.75, -2.25 \right],~\left[ -2.25, -1.75 \right],~\dots,~\left[ 2.25, 2.75 \right]$. Note, that all those values are successively normalized following \cref{tab:cartpole}, before being provided as initial observations to the agent. We evaluate the generalization over different initial conditions in terms of the \emph{attraction rate}, which quantifies the fraction of succeeded test runs. More formally, for a reward of $r=+1$ for each successful step, this reads:
\begin{align} \label{eq:rate_of_attraction}
    \frac{1}{\mathrm{runs}} \cdot \delta_{T=\sum_{i}^{T}r_i}
\end{align}
Hereby, $\delta$ is an indicator function, that evaluates to $1$, iff the run has achieved the maximal reward. In the considered setup, the horizon is $T=200$, and $100$ runs were performed for each setup.

The first thing one might notice about \cref{fig:generalization_regqpg} is that the results are not mirrored along the zero pole angular velocity axis. This is caused by negative velocities potentially offsetting large positive angles in the successive steps, while matters only get worse for positive velocities. However, the results are approximately point symmetrical (w.r.t the all-zero position) for negative angles, which we have skipped to avoid clustering of the plots. To simplify the interpretation of the results, we marked cells for which the regularized versions are more or less stable than the non-regularized baseline. We consider a setup to be \emph{considerable} better (or worse) if the standard deviations -- computed over all $100$ models -- do not overlap. If only the halved standard deviations do not overlap, we consider this to be a \emph{slightly} better (or worse) performance. 

A regularization rate of $\lambda=0.1$ improves the generalization compared to the non-regularized baseline on various initial conditions. However, the gain is especially pronounced for positive angular velocities. This trend is continued for $\lambda=0.2$, where even larger improvements for positive velocities can be reported, while the 
performance deteriorates slightly for negative velocities. This can be explained by the regularization during training limiting the impact of state input on the model output, which also includes the signs of values. For certain tasks, one might only want to \emph{regularize} certain properties of the state. This could be done e.g. by encoding the state magnitude and state sign separately, while only applying regularization to the parameters associated with the former one. However, such extensions are left for future work. For a high regularization rate of $\lambda=0.4$, apart from some exceptions, the trained models are less stable than the baseline. This is expected behavior, as too large regularization rates tend to obstruct the overall performance. Overall, depending on what regions the policy acts, a regularization rate between $\lambda=0.1$ and $\lambda=0.2$ is optimal for the considered setup.


\subsection{Incorporating Curriculum Learning}\label{subsec:generalization_cregqpg}

\begin{algorithm}[tb]
    \caption{Curriculum Reg. QPG (CurrRegQPG)}
    \label{alg:safe}
    \begin{algorithmic}
        \State {\bfseries Input:} policy approximator $\pi$, learning rate $\alpha$, regularization rate $\lambda$, maximum number of failures $f_{\max}$, increasing range of initial conditions $r_1, r_2, \dots, r_m$
        \State {\bfseries Output:} increasingly general policies $\pi^*_{r_1}, \pi^*_{r_2}, \dots$
        \State
        \State Initialize parameters, e.g.\ $\nu \sim [-\pi, \pi]$, $\omega \sim \mathcal{N}(0; 0.1)$
        \State Initialize failure counter $f \gets 0$ 
        \State Select first range of initial conditions $r \gets r_1$
        \While {Number of failures $f < f_{\max}$}
        \State Sample initial state $s_1$ from range $r$
        \State Follow policy $\pi_{\Theta=\lbrace\nu,\omega\rbrace}$: $s_1, a_1, r_1, \dots, s_T, a_T, r_T$ 
        \If {Episode lead to failure}
            \State Increase failure counter $f \gets f + 1$
        \EndIf
        \If {Validation generalizes on range $r$}
            \If {Reached final range $r = r_m$}
                \State Store $\pi^*_{r_m} \gets \pi_{\Theta}$ and terminate
            \EndIf
            \State Select next range $r ~(=r_i) \gets r_{i+1}$
            \State Store intermediate policy $\pi^*_{r_i} \gets \pi_{\Theta}$
        \EndIf
        \State Update $\Theta\gets\Theta+\alpha\nabla_{\Theta}J_{\mathrm{reg}(\lambda)}(\Theta)$, see Eq.~\ref{eq:reg_grad}
        \EndWhile
    \end{algorithmic}
\end{algorithm}

While \cref{subsec:generalization_regqpg} aimed at training policies that generalize well after training, we now  additionally want to minimize the number of failures during training. To this end, we propose the \acf{cregqpg} approach in \cref{alg:safe}, which closely resembles the concept of curriculum learning from classical \ac{rl}~\citep{elman1993learning,rohde1999language}. It incorporates three core concepts: (i) a standard quantum policy gradients approach, as proposed in~\cite{meyer2023quantum}; (ii) a regularization strategy based on Lipschitz bounds, as described in \cref{sec:training}, and analyzed regarding generalization in the previous paragraphs; (iii) an incrementally increasing range of initial conditions, intended to minimize the probability of failures, similar to~\cite{berkenkamp2017safe};
The algorithm initially trains on a small range of initial conditions, and keeps track of the number of failures -- i.e.\ validating conditions on either angel or spatial coordinates for \texttt{CartPole} -- during this process. Validation is performed to determine if the policy sufficiently generalizes over current range. If this is the case, the range of initial conditions is increased. This procedure continues until either the maximum number of failures has been reached, or generalization on the final range has been achieved. While this does not guarantee that no failures occur during training, the number can be significantly reduced in the presence of regularization, as we demonstrate in the following.

\begin{figure}[tbp]
    \centering
    \includegraphics{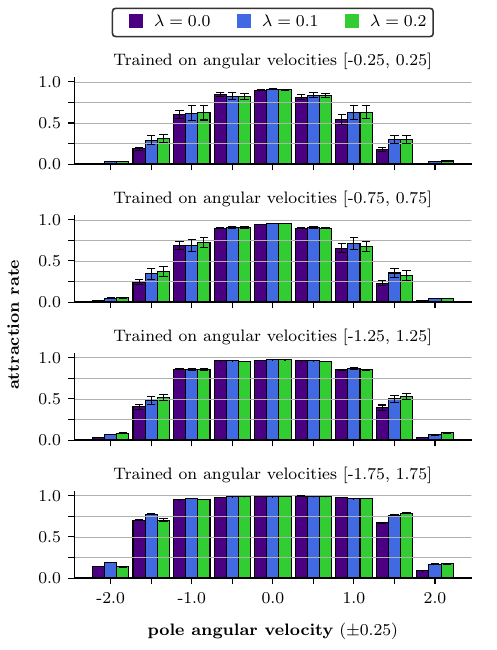}
    \caption{\label{fig:generalization_cregqpg}Generalization of policies for the models trained with \ac{cregqpg} on increasing ranges of initial conditions from \cref{tab:generalization}. The results are averaged only over the converged training runs, with the error bars denoting the standard deviations. The generalization was evaluated for $100$ test runs each, i.e. up to $10000$ samples contribute to each data point.}
\end{figure}

\begin{table}[tbp]
    \centering
    \caption{\label{tab:generalization}Lipschitz regularization minimizes number of failures during training of policies with the \ac{cregqpg} algorithm. We consider increasing ranges of initial conditions for the pole angular velocity, while all other state elements are initialized following the default procedure in \cref{tab:cartpole}. We report the average number of failures that occurred during training to ensure generalization over the respective range, with the maximum number of failures set to $1000$. Furthermore, we note the fraction of runs -- out of $100$ -- that was able to guarantee generalization over the respective configuration.}
    \begin{tabular*}{\linewidth}{@{\extracolsep{\fill}}c|cccc}
        \toprule
        Failures & $\lambda=0.0$ & $\lambda=0.1$ & $\lambda=0.2$ \\
        \midrule
        $~~\left[-0.25,~0.25\right]~$ & $208.9$ & $178.3$ & $\mathbf{167.9}$ \\
        $~~\left[-0.75,~0.75\right]~$ & $280.4$ & $\mathbf{228.3}$ & $\mathbf{228.3}$ \\
        $~~\left[-1.25,~1.25\right]~$ & $479.1$ & $\mathbf{394.6}$ & $489.0$ \\
        $~~\left[-1.75,~1.75\right]~$ & $990.8$ & $\mathbf{762.5}$ & $952.8$ \\
        \midrule
        \midrule
        Converged & $\lambda=0.0$ & $\lambda=0.1$ & $\lambda=0.2$ \\
        \midrule
        $~~\left[-0.25,~0.25\right]~$ & $99\%$ & $\mathbf{100\%}$ & $\mathbf{100\%}$ \\
        $~~\left[-0.75,~0.75\right]~$ & $97\%$ & $\mathbf{100\%}$ & $\mathbf{100\%}$ \\
        $~~\left[-1.25,~1.25\right]~$ & $83\%$ & $\mathbf{85\%}$ & $78\%$ \\
        $~~\left[-1.75,~1.75\right]~$ & $3\%$ & $\mathbf{41\%}$ & $8\%$ \\
        \bottomrule
    \end{tabular*}
\end{table}

For our proof-of-concept realization, we concentrated on incrementally enlarging the range of initial conditions for the pole angular velocity. We assert validation to be successful, if the average reward over $100$ runs exceeds $195$ -- i.e.\ the default termination condition for the \texttt{CartPole} environment. Moreover, we set a cut-off after at most $1000$ failures, and increased ranges from $[-0.25,0.25]$ to $[-1.75, 1.75]$ in four steps. We report the empirical results in \cref{tab:generalization}. The results indicate, that for a regularization rate of $\lambda=0.1$, the number of failures required for the policy to generalize over all ranges is reduced, compared to the non-regularized baseline. Moreover, especially the fraction of runs that were able to achieve generalization over the final range $[-1.75,1.75]$ significantly increases from $3\%$ to $41\%$ in the presence of regularization. For a larger regularization rate of $\lambda=0.2$, one can still observe a small advantage compared to the baseline on the larger ranges, but the difference is much less significant.

To provide more insights into the actually learned policies, we evaluated the generalization behavior at the intermediate steps in \cref{fig:generalization_cregqpg}. The plots only present data from the actually terminated runs, which leads to only small differences between the models. However, one can identify a similar pattern as in \cref{fig:generalization_regqpg}, i.e. the regularized models better generalize over initial conditions not explicitly trained on. Moreover, the worse generalization for $\lambda=0.2$ on negative vs. positive angular velocities explains the overall worse performance during cautious training. In general, the \ac{cregqpg} algorithm can enhance the generalization of policies on the desired ranges. A note regarding this, as especially for the bottom plot there are some bars with an attraction rate of only approx. $0.75$: validation in \ac{cregqpg} determines success as reaching an average reward of $195$ over $100$ runs, which means that several instances could have fallen short of reward $200$; generalization is evaluated as the fraction of runs that have reached reward $200$, i.e.\ a stricter condition; Overall, a regularization rate of $\lambda=0.1$ seems to be an optimal parameter for the \ac{cregqpg} algorithm in the considered setup.


\section{Discussion}\label{sec:discussion}

In this paper, we have explored the robustness and generalization capabilities of \acf{qrl} by integrating principles from quantum computing, classical \ac{rl}, and control theory. Our primary contribution is the development of the \acf{regqpg} algorithm. It leverages Lipschitz bounds to enhance the robustness and generalization of \ac{qrl} policies. Additionally, we proposed the \acf{cregqpg} approach to reduce the number of failures during training. Our numerical experiments demonstrated the effectiveness of the \ac{regqpg} and \ac{cregqpg} algorithms.

One of the key strengths of the proposed methods is their computational efficiency. Applying the regularization is computationally cheap, making it feasible to implement these strategies in various scenarios and applications. This efficiency is particularly important considering the scarcity of quantum computing resources, especially in the NISQ era.

Future work should consider less conservative conditions than those based on the Lipschitz bound in \cref{eq:lipschitz_bound}. While the Lipschitz bound provides a solid foundation for ensuring robustness, it may limit the performance of \ac{qrl} policies in certain scenarios. Exploring alternative conditions that balance robustness and performance more effectively could circumvent this. Furthermore, it might be interesting to investigate the integration of these regularization methods with \ac{qrl} algorithms beyond quantum policy gradients.

Finally, our work provides the basis for studying safety of \ac{qrl} algorithms.
In classical \ac{rl}, safety has been extensively studied~\citep{gu2022review}. Various methods have been proposed, such as safe \ac{rl} via Lyapunov methods and related approaches~\citep{berkenkamp2017safe,chow2018lyapunov,dawson2023safe}. In these approaches, guarantees are often provided by restricting the Lipschitz bound of the policy, as demonstrated in~\cite{berkenkamp2017safe,dawson2023safe}.
Lipschitz regularization in \ac{rl} has also been explored in~\cite{berkenkamp2017safe,bjorck2021towards,gogianu2021spectral,takase2022stability}, highlighting the important role that the Lipschitz bound plays.
Likewise, we expect that our framework paves the way for studying safety in \ac{qrl}.

In summary, this paper has made significant strides in enhancing the robustness and generalization capabilities of \ac{qrl} by incorporating Lipschitz bounds. Our findings underscore the importance of efficient regularization techniques and open new avenues for future research to expand the applicability of \ac{qrl} in the real world.


\section*{Code Availability}

Implementations of the \ac{regqpg} and \ac{cregqpg} algorithms are available in the repository \url{https://github.com/nicomeyer96/regularized-qpg}. The framework allows for full reproducibility of the experimental results in this paper by executing a single script. Usage instructions and additional details can be found in the \texttt{README} file. Further information and data is available upon reasonable request.


\section*{Acknowledgments}

NM acknowledges support by the Bavarian Ministry of Economic Affairs, Regional Development and Energy with funds from the Hightech Agenda Bayern via the project BayQS.
JB acknowledges funding by Deutsche Forschungsgemeinschaft (DFG, German Research Foundation) under Germany's Excellence Strategy - EXC 2075 - 390740016 and the support by the Stuttgart Center for Simulation Science (SimTech).


\bibliography{references}


\begin{thebibliography}{40}
\ifx \bisbn   \undefined \def \bisbn  #1{ISBN #1}\fi
\ifx \binits  \undefined \def \binits#1{#1}\fi
\ifx \bauthor  \undefined \def \bauthor#1{#1}\fi
\ifx \batitle  \undefined \def \batitle#1{#1}\fi
\ifx \bjtitle  \undefined \def \bjtitle#1{#1}\fi
\ifx \bvolume  \undefined \def \bvolume#1{\textbf{#1}}\fi
\ifx \byear  \undefined \def \byear#1{#1}\fi
\ifx \bissue  \undefined \def \bissue#1{#1}\fi
\ifx \bfpage  \undefined \def \bfpage#1{#1}\fi
\ifx \blpage  \undefined \def \blpage #1{#1}\fi
\ifx \burl  \undefined \def \burl#1{\textsf{#1}}\fi
\ifx \doiurl  \undefined \def \doiurl#1{\url{https://doi.org/#1}}\fi
\ifx \betal  \undefined \def \betal{\textit{et al.}}\fi
\ifx \binstitute  \undefined \def \binstitute#1{#1}\fi
\ifx \binstitutionaled  \undefined \def \binstitutionaled#1{#1}\fi
\ifx \bctitle  \undefined \def \bctitle#1{#1}\fi
\ifx \beditor  \undefined \def \beditor#1{#1}\fi
\ifx \bpublisher  \undefined \def \bpublisher#1{#1}\fi
\ifx \bbtitle  \undefined \def \bbtitle#1{#1}\fi
\ifx \bedition  \undefined \def \bedition#1{#1}\fi
\ifx \bseriesno  \undefined \def \bseriesno#1{#1}\fi
\ifx \blocation  \undefined \def \blocation#1{#1}\fi
\ifx \bsertitle  \undefined \def \bsertitle#1{#1}\fi
\ifx \bsnm \undefined \def \bsnm#1{#1}\fi
\ifx \bsuffix \undefined \def \bsuffix#1{#1}\fi
\ifx \bparticle \undefined \def \bparticle#1{#1}\fi
\ifx \barticle \undefined \def \barticle#1{#1}\fi
\bibcommenthead
\ifx \bconfdate \undefined \def \bconfdate #1{#1}\fi
\ifx \botherref \undefined \def \botherref #1{#1}\fi
\ifx \url \undefined \def \url#1{\textsf{#1}}\fi
\ifx \bchapter \undefined \def \bchapter#1{#1}\fi
\ifx \bbook \undefined \def \bbook#1{#1}\fi
\ifx \bcomment \undefined \def \bcomment#1{#1}\fi
\ifx \oauthor \undefined \def \oauthor#1{#1}\fi
\ifx \citeauthoryear \undefined \def \citeauthoryear#1{#1}\fi
\ifx \endbibitem  \undefined \def \endbibitem {}\fi
\ifx \bconflocation  \undefined \def \bconflocation#1{#1}\fi
\ifx \arxivurl  \undefined \def \arxivurl#1{\textsf{#1}}\fi
\csname PreBibitemsHook\endcsname

\bibitem[\protect\citeauthoryear{Brockman et~al.}{2016}]{brockman2016openai}
\begin{botherref}
\oauthor{\bsnm{Brockman}, \binits{G.}},
\oauthor{\bsnm{Cheung}, \binits{V.}},
\oauthor{\bsnm{Pettersson}, \binits{L.}},
\oauthor{\bsnm{Schneider}, \binits{J.}},
\oauthor{\bsnm{Schulman}, \binits{J.}},
\oauthor{\bsnm{Tang}, \binits{J.}},
\oauthor{\bsnm{Zaremba}, \binits{W.}}:
{O}pen{AI} {G}ym.
arXiv:1606.01540
(2016)
\end{botherref}
\endbibitem

\bibitem[\protect\citeauthoryear{Berberich et~al.}{2023}]{berberich2023training}
\begin{botherref}
\oauthor{\bsnm{Berberich}, \binits{J.}},
\oauthor{\bsnm{Fink}, \binits{D.}},
\oauthor{\bsnm{Pranji{\'c}}, \binits{D.}},
\oauthor{\bsnm{Tutschku}, \binits{C.}},
\oauthor{\bsnm{Holm}, \binits{C.}}:
Training robust and generalizable quantum models.
arXiv:2311.11871
(2023)
\end{botherref}
\endbibitem

\bibitem[\protect\citeauthoryear{Bjorck et~al.}{2021}]{bjorck2021towards}
\begin{bchapter}
\bauthor{\bsnm{Bjorck}, \binits{N.}},
\bauthor{\bsnm{Gomes}, \binits{C.P.}},
\bauthor{\bsnm{Weinberger}, \binits{K.Q.}}:
\bctitle{Towards deeper deep reinforcement learning with spectral normalization}.
In: \bbtitle{Advances in Neural Information Processing Systems},
vol. \bseriesno{34},
pp. \bfpage{8242}--\blpage{8255}
(\byear{2021})
\end{bchapter}
\endbibitem

\bibitem[\protect\citeauthoryear{Bravo-Prieto et~al.}{2023}]{bravo2023variational}
\begin{barticle}
\bauthor{\bsnm{Bravo-Prieto}, \binits{C.}},
\bauthor{\bsnm{LaRose}, \binits{R.}},
\bauthor{\bsnm{Cerezo}, \binits{M.}},
\bauthor{\bsnm{Subasi}, \binits{Y.}},
\bauthor{\bsnm{Cincio}, \binits{L.}},
\bauthor{\bsnm{Coles}, \binits{P.J.}}:
\batitle{Variational quantum linear solver}.
\bjtitle{Quantum}
\bvolume{7},
\bfpage{1188}
(\byear{2023})
\end{barticle}
\endbibitem

\bibitem[\protect\citeauthoryear{Berkenkamp et~al.}{2017}]{berkenkamp2017safe}
\begin{bchapter}
\bauthor{\bsnm{Berkenkamp}, \binits{F.}},
\bauthor{\bsnm{Turchetta}, \binits{M.}},
\bauthor{\bsnm{Schoellig}, \binits{A.}},
\bauthor{\bsnm{Krause}, \binits{A.}}:
\bctitle{Safe model-based reinforcement learning with stability guarantees}.
In: \bbtitle{Advances in Neural Information Processing Systems},
pp. \bfpage{908}--\blpage{918}
(\byear{2017})
\end{bchapter}
\endbibitem

\bibitem[\protect\citeauthoryear{Biamonte et~al.}{2017}]{biamonte2017quantum}
\begin{barticle}
\bauthor{\bsnm{Biamonte}, \binits{J.}},
\bauthor{\bsnm{Wittek}, \binits{P.}},
\bauthor{\bsnm{Pancotti}, \binits{N.}},
\bauthor{\bsnm{Rebentrost}, \binits{P.}},
\bauthor{\bsnm{Wiebe}, \binits{N.}},
\bauthor{\bsnm{Lloyd}, \binits{S.}}:
\batitle{Quantum machine learning}.
\bjtitle{Nature}
\bvolume{549}(\bissue{7671}),
\bfpage{195}--\blpage{202}
(\byear{2017})
\end{barticle}
\endbibitem

\bibitem[\protect\citeauthoryear{Cerezo et~al.}{2021}]{cerezo2021variational}
\begin{barticle}
\bauthor{\bsnm{Cerezo}, \binits{M.}},
\bauthor{\bsnm{Arrasmith}, \binits{A.}},
\bauthor{\bsnm{Babbush}, \binits{R.}},
\bauthor{\bsnm{Benjamin}, \binits{S.C.}},
\bauthor{\bsnm{Endo}, \binits{S.}},
\bauthor{\bsnm{Fujii}, \binits{K.}},
\bauthor{\bsnm{McClean}, \binits{J.R.}},
\bauthor{\bsnm{Mitarai}, \binits{K.}},
\bauthor{\bsnm{Yuan}, \binits{X.}},
\bauthor{\bsnm{Cincio}, \binits{L.}},
\bauthor{\bsnm{Coles}, \binits{P.J.}}:
\batitle{Variational quantum algorithms}.
\bjtitle{Nat. Rev. Phys.}
\bvolume{3},
\bfpage{625}--\blpage{644}
(\byear{2021})
\end{barticle}
\endbibitem

\bibitem[\protect\citeauthoryear{Chow et~al.}{2018}]{chow2018lyapunov}
\begin{bchapter}
\bauthor{\bsnm{Chow}, \binits{Y.}},
\bauthor{\bsnm{Nachum}, \binits{O.}},
\bauthor{\bsnm{Duenez-Guzman}, \binits{E.}},
\bauthor{\bsnm{Ghavamzadeh}, \binits{M.}}:
\bctitle{A lyapunov-based approach to safe reinforcement learning}.
In: \bbtitle{Advances in Neural Information Processing Systems},
vol. \bseriesno{31}
(\byear{2018})
\end{bchapter}
\endbibitem

\bibitem[\protect\citeauthoryear{Chen et~al.}{2020}]{chen2020variational}
\begin{barticle}
\bauthor{\bsnm{Chen}, \binits{S.Y.-C.}},
\bauthor{\bsnm{Yang}, \binits{C.-H.H.}},
\bauthor{\bsnm{Qi}, \binits{J.}},
\bauthor{\bsnm{Chen}, \binits{P.-Y.}},
\bauthor{\bsnm{Ma}, \binits{X.}},
\bauthor{\bsnm{Goan}, \binits{H.-S.}}:
\batitle{Variational quantum circuits for deep reinforcement learning}.
\bjtitle{IEEE Access}
\bvolume{8},
\bfpage{141007}--\blpage{141024}
(\byear{2020})
\end{barticle}
\endbibitem

\bibitem[\protect\citeauthoryear{Dawson et~al.}{2023}]{dawson2023safe}
\begin{barticle}
\bauthor{\bsnm{Dawson}, \binits{C.}},
\bauthor{\bsnm{Gao}, \binits{S.}},
\bauthor{\bsnm{Fan}, \binits{C.}}:
\batitle{Safe control with learned certificates: A survey of neural lyapunov, barrier, and contraction methods for robotics and control}.
\bjtitle{IEEE Trans. Robotics}
\bvolume{39}(\bissue{3}),
\bfpage{1749}--\blpage{1767}
(\byear{2023})
\end{barticle}
\endbibitem

\bibitem[\protect\citeauthoryear{Elman}{1993}]{elman1993learning}
\begin{barticle}
\bauthor{\bsnm{Elman}, \binits{J.L.}}:
\batitle{Learning and development in neural networks: The importance of starting small}.
\bjtitle{Cognition}
\bvolume{48}(\bissue{1}),
\bfpage{71}--\blpage{99}
(\byear{1993})
\end{barticle}
\endbibitem

\bibitem[\protect\citeauthoryear{Gogianu et~al.}{2021}]{gogianu2021spectral}
\begin{bchapter}
\bauthor{\bsnm{Gogianu}, \binits{F.}},
\bauthor{\bsnm{Berariu}, \binits{T.}},
\bauthor{\bsnm{Rosca}, \binits{M.C.}},
\bauthor{\bsnm{Clopath}, \binits{C.}},
\bauthor{\bsnm{Busoniu}, \binits{L.}},
\bauthor{\bsnm{Pascanu}, \binits{R.}}:
\bctitle{Spectral normalisation for deep reinforcement learning: An optimisation perspective}.
In: \beditor{\bsnm{Meila}, \binits{M.}},
\beditor{\bsnm{Zhang}, \binits{T.}} (eds.)
\bbtitle{Proc. 38th Int. Conf. Machine Learning (ICML)},
vol. \bseriesno{139},
pp. \bfpage{3734}--\blpage{3744}
(\byear{2021})
\end{bchapter}
\endbibitem

\bibitem[\protect\citeauthoryear{Goodfellow et~al.}{2014}]{goodfellow2014explaining}
\begin{botherref}
\oauthor{\bsnm{Goodfellow}, \binits{I.J.}},
\oauthor{\bsnm{Shlens}, \binits{J.}},
\oauthor{\bsnm{Szegedy}, \binits{C.}}:
Explaining and harnessing adversarial examples.
{arXiv:1412.6572}
(2014)
\end{botherref}
\endbibitem

\bibitem[\protect\citeauthoryear{Gu et~al.}{2022}]{gu2022review}
\begin{botherref}
\oauthor{\bsnm{Gu}, \binits{S.}},
\oauthor{\bsnm{Yang}, \binits{L.}},
\oauthor{\bsnm{Du}, \binits{Y.}},
\oauthor{\bsnm{Chen}, \binits{G.}},
\oauthor{\bsnm{Walter}, \binits{F.}},
\oauthor{\bsnm{Wang}, \binits{J.}},
\oauthor{\bsnm{Knoll}, \binits{A.}}:
A review of safe reinforcement learning: Methods, theory and applications.
arXiv:2205.10330
(2022)
\end{botherref}
\endbibitem

\bibitem[\protect\citeauthoryear{Javadi-Abhari et~al.}{2024}]{qiskit2024}
\begin{botherref}
\oauthor{\bsnm{Javadi-Abhari}, \binits{A.}},
\oauthor{\bsnm{Treinish}, \binits{M.}},
\oauthor{\bsnm{Krsulich}, \binits{K.}},
\oauthor{\bsnm{Wood}, \binits{C.J.}},
\oauthor{\bsnm{Lishman}, \binits{J.}},
\oauthor{\bsnm{Gacon}, \binits{J.}},
\oauthor{\bsnm{Martiel}, \binits{S.}},
\oauthor{\bsnm{Nation}, \binits{P.D.}},
\oauthor{\bsnm{Bishop}, \binits{L.S.}},
\oauthor{\bsnm{Cross}, \binits{A.W.}},
\oauthor{\bsnm{Johnson}, \binits{B.R.}},
\oauthor{\bsnm{Gambetta}, \binits{J.M.}}:
Quantum computing with {Q}iskit
(2024)
\end{botherref}
\endbibitem

\bibitem[\protect\citeauthoryear{Jerbi et~al.}{2021}]{jerbi2021parametrized}
\begin{bchapter}
\bauthor{\bsnm{Jerbi}, \binits{S.}},
\bauthor{\bsnm{Gyurik}, \binits{C.}},
\bauthor{\bsnm{Marshall}, \binits{S.}},
\bauthor{\bsnm{Briegel}, \binits{H.}},
\bauthor{\bsnm{Dunjko}, \binits{V.}}:
\bctitle{Parametrized quantum policies for reinforcement learning}.
In: \bbtitle{Advances in Neural Information Processing Systems},
vol. \bseriesno{34},
pp. \bfpage{28362}--\blpage{28375}
(\byear{2021})
\end{bchapter}
\endbibitem

\bibitem[\protect\citeauthoryear{Liu et~al.}{2021}]{liu2021rigorous}
\begin{barticle}
\bauthor{\bsnm{Liu}, \binits{Y.}},
\bauthor{\bsnm{Arunachalam}, \binits{S.}},
\bauthor{\bsnm{Temme}, \binits{K.}}:
\batitle{A rigorous and robust quantum speed-up in supervised machine learning}.
\bjtitle{Nature Physics}
\bvolume{17}(\bissue{9}),
\bfpage{1013}--\blpage{1017}
(\byear{2021})
\end{barticle}
\endbibitem

\bibitem[\protect\citeauthoryear{Meyer et~al.}{2024}]{meyer2024warm}
\begin{botherref}
\oauthor{\bsnm{Meyer}, \binits{N.}},
\oauthor{\bsnm{Murauer}, \binits{J.}},
\oauthor{\bsnm{Popov}, \binits{A.}},
\oauthor{\bsnm{Ufrecht}, \binits{C.}},
\oauthor{\bsnm{Plinge}, \binits{A.}},
\oauthor{\bsnm{Mutschler}, \binits{C.}},
\oauthor{\bsnm{Scherer}, \binits{D.D.}}:
Warm-start variational quantum policy iteration.
arXiv preprint arXiv:2404.10546
(2024)
\end{botherref}
\endbibitem

\bibitem[\protect\citeauthoryear{Meyer et~al.}{2024}]{meyer2024comprehensive}
\begin{botherref}
\oauthor{\bsnm{Meyer}, \binits{N.}},
\oauthor{\bsnm{R{\"o}hn}, \binits{M.}},
\oauthor{\bsnm{Murauer}, \binits{J.}},
\oauthor{\bsnm{Plinge}, \binits{A.}},
\oauthor{\bsnm{Mutschler}, \binits{C.}},
\oauthor{\bsnm{Scherer}, \binits{D.D.}}:
Comprehensive library of variational lse solvers.
arXiv preprint arXiv:2404.09916
(2024)
\end{botherref}
\endbibitem

\bibitem[\protect\citeauthoryear{Meyer et~al.}{2023a}]{meyer2023quantum}
\begin{bchapter}
\bauthor{\bsnm{Meyer}, \binits{N.}},
\bauthor{\bsnm{Scherer}, \binits{D.D.}},
\bauthor{\bsnm{Plinge}, \binits{A.}},
\bauthor{\bsnm{Mutschler}, \binits{C.}},
\bauthor{\bsnm{Hartmann}, \binits{M.J.}}:
\bctitle{Quantum policy gradient algorithm with optimized action decoding}.
In: \bbtitle{Proc. 40th Int. Conf. Machine Learning (ICML)},
vol. \bseriesno{202},
pp. \bfpage{24592}--\blpage{24613}
(\byear{2023})
\end{bchapter}
\endbibitem

\bibitem[\protect\citeauthoryear{Meyer et~al.}{2023b}]{meyer2023natural}
\begin{bchapter}
\bauthor{\bsnm{Meyer}, \binits{N.}},
\bauthor{\bsnm{Scherer}, \binits{D.D.}},
\bauthor{\bsnm{Plinge}, \binits{A.}},
\bauthor{\bsnm{Mutschler}, \binits{C.}},
\bauthor{\bsnm{Hartmann}, \binits{M.J.}}:
\bctitle{Quantum natural policy gradients: Towards sample-efficient reinforcement learning}.
In: \bbtitle{2023 IEEE International Conference on Quantum Computing and Engineering (QCE)},
vol. \bseriesno{2},
pp. \bfpage{36}--\blpage{41}
(\byear{2023})
\end{bchapter}
\endbibitem

\bibitem[\protect\citeauthoryear{Meyer et~al.}{2022}]{meyer2022survey}
\begin{botherref}
\oauthor{\bsnm{Meyer}, \binits{N.}},
\oauthor{\bsnm{Ufrecht}, \binits{C.}},
\oauthor{\bsnm{Periyasamy}, \binits{M.}},
\oauthor{\bsnm{Scherer}, \binits{D.D.}},
\oauthor{\bsnm{Plinge}, \binits{A.}},
\oauthor{\bsnm{Mutschler}, \binits{C.}}:
A survey on quantum reinforcement learning.
arXiv:2211.03464
(2022)
\end{botherref}
\endbibitem

\bibitem[\protect\citeauthoryear{Meyer et~al.}{2024}]{meyer2024qiskit}
\begin{botherref}
\oauthor{\bsnm{Meyer}, \binits{N.}},
\oauthor{\bsnm{Ufrecht}, \binits{C.}},
\oauthor{\bsnm{Periyasamy}, \binits{M.}},
\oauthor{\bsnm{Plinge}, \binits{A.}},
\oauthor{\bsnm{Mutschler}, \binits{C.}},
\oauthor{\bsnm{Scherer}, \binits{D.D.}},
\oauthor{\bsnm{Maier}, \binits{A.}}:
Qiskit-torch-module: Fast prototyping of quantum neural networks.
arXiv:2404.06314
(2024)
\end{botherref}
\endbibitem

\bibitem[\protect\citeauthoryear{Nielsen and Chuang}{2011}]{nielsen2011quantum}
\begin{bbook}
\bauthor{\bsnm{Nielsen}, \binits{M.A.}},
\bauthor{\bsnm{Chuang}, \binits{I.L.}}:
\bbtitle{Quantum Computation and Quantum Information: 10th Anniversary Edition},
\bedition{10th ed.} edn.
\bpublisher{Cambridge University Press},
\blocation{New York, NY, USA}
(\byear{2011})
\end{bbook}
\endbibitem

\bibitem[\protect\citeauthoryear{Packer et~al.}{2019}]{packer2019assessing}
\begin{botherref}
\oauthor{\bsnm{Packer}, \binits{C.}},
\oauthor{\bsnm{Gao}, \binits{K.}},
\oauthor{\bsnm{Kos}, \binits{J.}},
\oauthor{\bsnm{Kr{\"a}henb{\"u}hl}, \binits{P.}},
\oauthor{\bsnm{Koltun}, \binits{V.}},
\oauthor{\bsnm{Song}, \binits{D.}}:
Assessing generalization in deep reinforcement learning.
arXiv:1810.12282
(2019)
\end{botherref}
\endbibitem

\bibitem[\protect\citeauthoryear{Periyasamy et~al.}{2023}]{periyasamy2023bcqq}
\begin{botherref}
\oauthor{\bsnm{Periyasamy}, \binits{M.}},
\oauthor{\bsnm{H{\"o}lle}, \binits{M.}},
\oauthor{\bsnm{Wiedmann}, \binits{M.}},
\oauthor{\bsnm{Scherer}, \binits{D.D.}},
\oauthor{\bsnm{Plinge}, \binits{A.}},
\oauthor{\bsnm{Mutschler}, \binits{C.}}:
Bcqq: Batch-constraint quantum q-learning with cyclic data re-uploading.
arXiv preprint arXiv:2305.00905
(2023)
\end{botherref}
\endbibitem

\bibitem[\protect\citeauthoryear{Preskill}{2018}]{preskill2018quantum}
\begin{barticle}
\bauthor{\bsnm{Preskill}, \binits{J.}}:
\batitle{Quantum computing in the {NISQ} era and beyond}.
\bjtitle{Quantum}
\bvolume{2},
\bfpage{79}
(\byear{2018})
\end{barticle}
\endbibitem

\bibitem[\protect\citeauthoryear{Rohde and Plaut}{1999}]{rohde1999language}
\begin{barticle}
\bauthor{\bsnm{Rohde}, \binits{D.L.}},
\bauthor{\bsnm{Plaut}, \binits{D.C.}}:
\batitle{Language acquisition in the absence of explicit negative evidence: How important is starting small?}
\bjtitle{Cognition}
\bvolume{72}(\bissue{1}),
\bfpage{67}--\blpage{109}
(\byear{1999})
\end{barticle}
\endbibitem

\bibitem[\protect\citeauthoryear{Sutton and Barto}{2018}]{sutton2018reinforcement}
\begin{bbook}
\bauthor{\bsnm{Sutton}, \binits{R.}},
\bauthor{\bsnm{Barto}, \binits{A.}}:
\bbtitle{Reinforcement Learning: An Introduction}.
\bpublisher{MIT Press},
\blocation{Cambridge, MA, USA}
(\byear{2018})
\end{bbook}
\endbibitem

\bibitem[\protect\citeauthoryear{Skolik et~al.}{2022}]{skolik2022quantum}
\begin{barticle}
\bauthor{\bsnm{Skolik}, \binits{A.}},
\bauthor{\bsnm{Jerbi}, \binits{S.}},
\bauthor{\bsnm{Dunjko}, \binits{V.}}:
\batitle{Quantum agents in the {Gym}: a variational quantum algorithm for deep {Q}-learning}.
\bjtitle{Quantum}
\bvolume{6},
\bfpage{720}
(\byear{2022})
\end{barticle}
\endbibitem

\bibitem[\protect\citeauthoryear{Sutton et~al.}{1999}]{sutton1999policy}
\begin{botherref}
\oauthor{\bsnm{Sutton}, \binits{R.S.}},
\oauthor{\bsnm{McAllester}, \binits{D.}},
\oauthor{\bsnm{Singh}, \binits{S.}},
\oauthor{\bsnm{Mansour}, \binits{Y.}}:
Policy gradient methods for reinforcement learning with function approximation.
Advances in neural information processing systems
\textbf{12}
(1999)
\end{botherref}
\endbibitem

\bibitem[\protect\citeauthoryear{Szegedy et~al.}{2014}]{szegedy2014intriguing}
\begin{botherref}
\oauthor{\bsnm{Szegedy}, \binits{C.}},
\oauthor{\bsnm{Zaremba}, \binits{W.}},
\oauthor{\bsnm{Sutskever}, \binits{I.}},
\oauthor{\bsnm{Bruna}, \binits{J.}},
\oauthor{\bsnm{Erhan}, \binits{D.}},
\oauthor{\bsnm{Goodfellow}, \binits{I.}},
\oauthor{\bsnm{Fergus}, \binits{R.}}:
Intriguing properties of neural networks.
{arXiv:1312.6199}
(2014)
\end{botherref}
\endbibitem

\bibitem[\protect\citeauthoryear{Tamar et~al.}{2015}]{tamar2015optimizing}
\begin{bchapter}
\bauthor{\bsnm{Tamar}, \binits{A.}},
\bauthor{\bsnm{Glassner}, \binits{Y.}},
\bauthor{\bsnm{Mannor}, \binits{S.}}:
\bctitle{Optimizing the {CVaR} via sampling}.
In: \bbtitle{Proc. AAAI Conf. Artificial Intelligence (AAAI)}
(\byear{2015})
\end{bchapter}
\endbibitem

\bibitem[\protect\citeauthoryear{Takase et~al.}{2022}]{takase2022stability}
\begin{barticle}
\bauthor{\bsnm{Takase}, \binits{R.}},
\bauthor{\bsnm{Yoshikawa}, \binits{N.}},
\bauthor{\bsnm{Mariyama}, \binits{T.}},
\bauthor{\bsnm{Tsuchiya}, \binits{T.}}:
\batitle{Stability-certified reinforcement learning control via spectral normalization}.
\bjtitle{Mach. Learn. Appl.}
\bvolume{10},
\bfpage{100409}
(\byear{2022})
\end{barticle}
\endbibitem

\bibitem[\protect\citeauthoryear{Wu et~al.}{2020}]{wu2020quantum}
\begin{botherref}
\oauthor{\bsnm{Wu}, \binits{S.}},
\oauthor{\bsnm{Jin}, \binits{S.}},
\oauthor{\bsnm{Wen}, \binits{D.}},
\oauthor{\bsnm{Han}, \binits{D.}},
\oauthor{\bsnm{Wang}, \binits{X.}}:
Quantum reinforcement learning in continuous action space.
arXiv preprint arXiv:2012.10711
(2020)
\end{botherref}
\endbibitem

\bibitem[\protect\citeauthoryear{Wang et~al.}{2019}]{wang2019generalization}
\begin{bchapter}
\bauthor{\bsnm{Wang}, \binits{H.}},
\bauthor{\bsnm{Zheng}, \binits{S.}},
\bauthor{\bsnm{Xiong}, \binits{C.}},
\bauthor{\bsnm{Socher}, \binits{R.}}:
\bctitle{On the generalization gap in reparameterizable reinforcement learning}.
In: \bbtitle{Proc. Int. Conf. Machine Learning (ICML)},
vol. \bseriesno{97}.
\bpublisher{PMLR}, \blocation{???}
(\byear{2019})
\end{bchapter}
\endbibitem

\bibitem[\protect\citeauthoryear{Xu and Mannor}{2012}]{xu2012robustness}
\begin{barticle}
\bauthor{\bsnm{Xu}, \binits{H.}},
\bauthor{\bsnm{Mannor}, \binits{S.}}:
\batitle{Robustness and generalization}.
\bjtitle{Mach. Learn.}
\bvolume{86}(\bissue{3}),
\bfpage{391}--\blpage{423}
(\byear{2012})
\doiurl{10.1007/s10994-011-5268-1}
\end{barticle}
\endbibitem

\bibitem[\protect\citeauthoryear{Yun et~al.}{2022}]{yun2022quantum}
\begin{bchapter}
\bauthor{\bsnm{Yun}, \binits{W.J.}},
\bauthor{\bsnm{Kwak}, \binits{Y.}},
\bauthor{\bsnm{Kim}, \binits{J.P.}},
\bauthor{\bsnm{Cho}, \binits{H.}},
\bauthor{\bsnm{Jung}, \binits{S.}},
\bauthor{\bsnm{Park}, \binits{J.}},
\bauthor{\bsnm{Kim}, \binits{J.}}:
\bctitle{Quantum multi-agent reinforcement learning via variational quantum circuit design}.
In: \bbtitle{2022 IEEE 42nd International Conference on Distributed Computing Systems (ICDCS)},
pp. \bfpage{1332}--\blpage{1335}
(\byear{2022})
\end{bchapter}
\endbibitem

\bibitem[\protect\citeauthoryear{Yu et~al.}{2017}]{yu2017preparing}
\begin{botherref}
\oauthor{\bsnm{Yu}, \binits{W.}},
\oauthor{\bsnm{Tan}, \binits{J.}},
\oauthor{\bsnm{Liu}, \binits{C.K.}},
\oauthor{\bsnm{Turk}, \binits{G.}}:
Preparing for the unknown: learning a universal policy with online system identification.
Robotics: Science and Systems (RSS)
(2017)
\end{botherref}
\endbibitem

\bibitem[\protect\citeauthoryear{Zaman et~al.}{2023}]{zaman2023survey}
\begin{botherref}
\oauthor{\bsnm{Zaman}, \binits{K.}},
\oauthor{\bsnm{Marchisio}, \binits{A.}},
\oauthor{\bsnm{Hanif}, \binits{M.A.}},
\oauthor{\bsnm{Shafique}, \binits{M.}}:
A survey on quantum machine learning: Current trends, challenges, opportunities, and the road ahead.
arXiv preprint arXiv:2310.10315
(2023)
\end{botherref}
\endbibitem

\end{thebibliography}

\end{document}